\documentclass[12pt,preprint]{aastex}
\usepackage{graphicx}

\begin{document}

\title{Depletion of CCS in a Candidate Warm-Carbon-Chain-Chemistry Source L483}
\author{Tomoya HIROTA}
\affil {National Astronomical Observatory of Japan, Mitaka, Tokyo 181-8588, Japan }
\affil {tomoya.hirota@nao.ac.jp}
\author{Nami SAKAI, and Satoshi YAMAMOTO}
\affil {Department of Physics, The University of Tokyo, Bunkyo-ku, Tokyo 113-0033, Japan }

\begin{abstract}
We have carried out an observation of the CCS ($J_{N}$=2$_{1}-$1$_{0}$) line 
with the Very Large Array in its D-configuration 
toward a protostellar core L483 (IRAS~18140$-$0440). 
This is a candidate source of the newly found carbon-chain rich environment called 
"Warm-Carbon-Chain-Chemistry (WCCC)", according to the previous observations of 
carbon-chain molecules. The CCS distribution in L483 is found to consist of 
two clumps aligned in the northwest-southeast 
direction, well tracing the CCS ridge observed with the single-dish radio telescope. 
The most remarkable feature is that CCS is depleted at the core center. 
Such a CCS distribution with the central hole is consistent with those of previously 
observed prestellar and protostellar cores, but it is rather unexpected for L483. 
This is because the distribution of CS, which is usually similar to that of CCS, 
is centrally peaked. 
Our results imply that the CCS ($J_{N}$=2$_{1}-$1$_{0}$) line would selectively trace 
the outer cold envelope in the chemically less evolved phase that is 
seriously resolved out with the interferometric observation. 
Thus, it is most likely that 
the high abundance of CCS in L483 relative to the other WCCC sources is not 
due to the activity of the protostar, although it would be related to its younger 
chemical evolutionary stage, or a short timescale of the prestellar phase. 
\end{abstract}

\keywords{ISM: individual objects (L483) --- ISM: molecules --- radio lines: ISM}

\section{Introduction}
We recently discovered a Class 0 protostellar core, L1527 in the Taurus molecular cloud, 
which is extremely rich in carbon-chain molecules \citep{sakai2008}. 
Based on the detailed observations, carbon-chain molecules and their anions, 
C$_{n}$H, C$_{n}$H$_{2}$, HC$_{n}$N, and C$_{n}$H$^{-}$, are confirmed to be enhanced in 
the warm and dense gas near the protostar, IRAS~04368+2557, in L1527. 
We call such a newly found chemical environment "Warm-Carbon-Chain-Chemistry (WCCC)". 
The origin of WCCC is proposed to be a consequence of 
regeneration of carbon-chain molecules triggered by evaporation of CH$_{4}$ 
from grain mantles in the warm ($\sim$30~K) gas around the protostar 
\citep{sakai2008, hassel2008}. 
This mechanism has never been expected in 
classical chemical models for cold dark clouds where carbon-chain molecules are 
efficiently produced mainly through ion-molecular reactions initiated by 
atomic carbon before being fixed into CO \citep{suzuki1992, bergin1997, aikawa2001}. 
Subsequently, we carried out a survey of several carbon-chain molecules toward 
low-mass protostellar cores, and found the second WCCC source, IRAS~15398$-$3359 
in the Lupus molecular cloud \citep{sakai2009}. 

In addition to L1527 and IRAS~15398$-$3359, some possible candidates for WCCC were recognized. 
One of them is L483, where a rare molecule, l-C$_{3}$H$_{2}$, 
is detected and the C$_{4}$H abundance is relatively high \citep{sakai2009}. 
L483 is a well studied star-forming core associated with a protostar 
IRAS~18140$-$0440 in the Aquila rift at a distance of 200~pc \citep{tafalla2000}. 
The evolutionary class of IRAS~18140$-$0440 (a transition from Class 0 to Class I) 
is almost similar to that of IRAS~04368+2557 in L1527. 
This source also seems to be outstanding in a survey of C$_{3}$H$_{2}$ 
\citep{benson1998}, in which the C$_{3}$H$_{2}$ line is bright next to 
TMC-1 (cyanopolyyne peak and NH$_{3}$ peak) and L1527. 

At the same time, we also identified L483 as a distinctly rich source of 
carbon-chain molecules like CCS, HC$_{3}$N, and HC$_{5}$N among 
star-forming dense cores \citep{hirota2009}. 
The CCS ($J_{N}$=4$_{3}-$3$_{2}$) line is much brighter toward L483 than 
toward the two other WCCC sources 
\citep[][N. Sakai et al. in preparation]{benson1998, hirota2009}. 
Mapping observations with the 45~m telescope at the Nobeyama Radio Observatory (NRO) 
revealed that the distribution of CCS has a centrally peaked structure. 
This seems consistent with the interferometric map of the CS ($J$=2$-$1) line 
which also shows a centrally peaked structure \citep{jorgensen2004}. 
Although these features of CCS could be related to WCCC, 
no signature of the CCS depletion may be due to a coarse spatial resolution (37\arcsec) of the NRO 45~m 
telescope \citep{hirota2009}. 
The resolution may be insufficient to resolve a central hole with 
the predicted radius of $<$5000~AU or 25\arcsec \ at the distance of L483 \citep{aikawa2001}. 
In fact, the central hole of CCS in a Class 0 protostellar core B335 is detected only 
by the interferometric observations \citep{velusamy1995}. 
With this in mind, we have conducted a mapping observation of 
the CCS line toward L483 using the Very Large Array (VLA) of 
the National Radio Astronomy Observatory (NRAO). 

\section{Observation and Data Analysis}

The observation was made with the VLA in its 
D-configuration on 2009 October 26. 
The observed line was CCS ($J_{N}$=2$_{1}-$1$_{0}$) 
whose rest frequency is 22344.033~MHz \citep{yamamoto1990}. 
Two intermediate frequency (IF) bands were employed to receive dual 
circular polarizations with the bandwidth of 3.125~MHz for each. 
A spectral resolution was set to be 24.414~kHz, or the velocity resolution of 0.32~km~s$^{-1}$. 
The target source L483, R.A.=18$^{\rm{h}}$17$^{\rm{m}}$29$^{\rm{s}}$.8 and 
decl. =$-$04$^{\rm{d}}$39\arcmin38\arcsec.3, and 
a secondary calibrator, 1743-038, which is 8$^{\circ}$.4 away from L483, were observed 
over a cycle time of 12~minutes. 
For each cycle, 9~minutes were spent on L483 and 3~minutes on 1743-038. 
The observation was allocated for 9~hr and the total on-source 
integration time was 4.2~hr. 

The data were edited and calibrated in the standard manner by 
using the NRAO Astronomical Image Processing System (AIPS) software package. 
The absolute flux density and bandpass calibrator was 1331+305 (3C286), 
for which we adopted a flux density of 2.53~Jy. Amplitudes and phases were 
calibrated by observing the secondary calibrator 1743-038. 
The bootstrapped flux density of 1743-038 was 2.99~Jy. 
The Doppler correction was applied in the post-processing using the AIPS task CVEL. 
Synthesis imaging and CLEAN were made by using the AIPS task IMAGR. 
In order to achieve higher sensitivity, we adopted the ROBUST weighting parameter 
of 5 (natural weighting) and the UVRANGE parameter from 0 to 25~k$\lambda$. 
The synthesized beam size was 7\arcsec.87$\times$7\arcsec.15 with the position 
angle of -28$^{\circ}$.68. 
The resulting image sensitivity was 4~mJy~beam$^{-1}$ for each spectral channel. 

\section{Results}

\begin{figure}[htb]
\begin{minipage}[htb]{8cm}
\epsscale{0.8}
\plotone{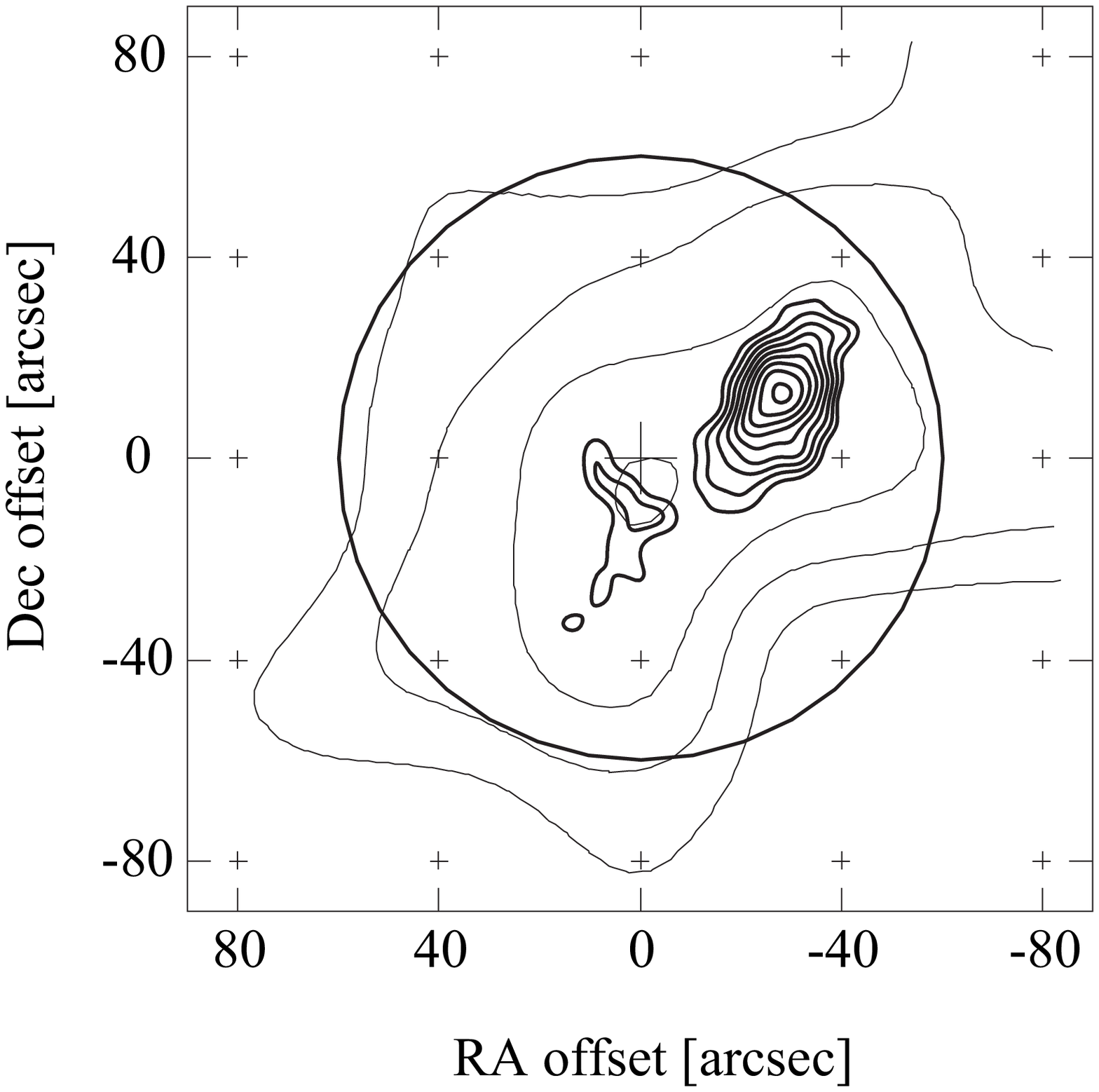}
\figcaption[]{Integrated intensity map of the CCS ($J_{N}$=2$_{1}-$1$_{0}$) line 
observed with the VLA (bold line). 
The velocity range for integration is 5.1 and 5.4~km~s$^{-1}$, and 
the interval of the contours for the VLA map is 6~mJy~beam$^{-1}$ (1$\sigma$) with 
the lowest one of 18~mJy~beam$^{-1}$ (3$\sigma$). A large cross indicates the radio 
counterpart of IRAS~18148-0440 \citep[VLA-2;][]{beltran2001}. 
A circle indicates the primary beam size of the VLA, and the data outside of this region 
are clipped for clarify the image. Thin contours represent the 
integrated intensity map of the CCS ($J_{N}$=4$_{3}-$3$_{2}$) line observed 
with the NRO 45~m telescope \citep{hirota2009}. 
The observed grid points are indicated by small crosses. 
\label{fig-map}}
\end{minipage}
\hspace{1.5cm}
\begin{minipage}[thb]{8cm}
\epsscale{0.8}
\plotone{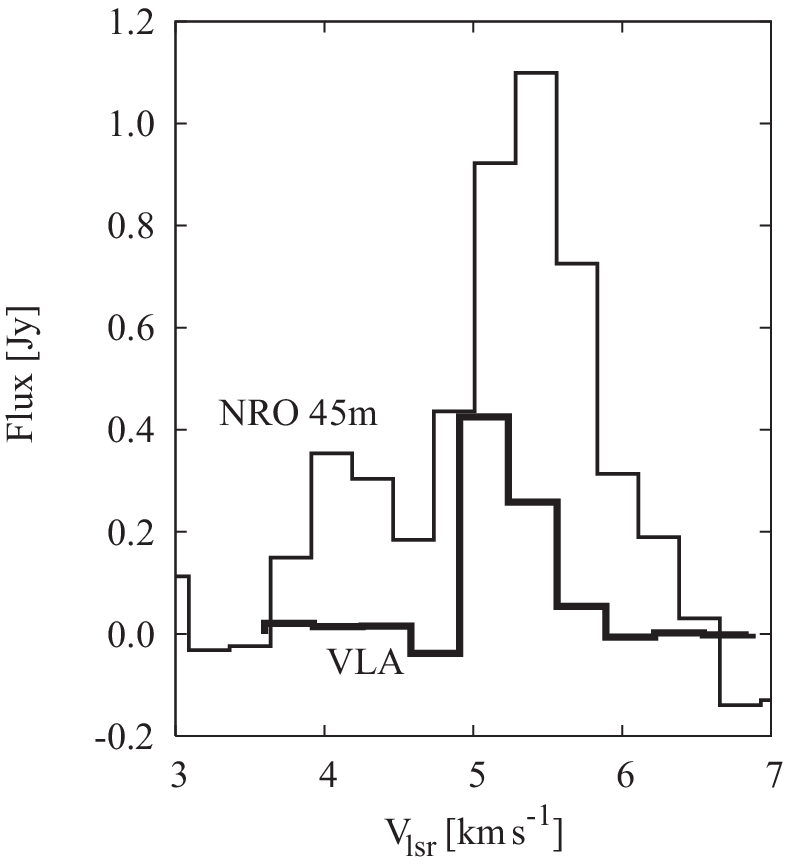}
\figcaption[]{Spectra of the CCS ($J_{N}$=2$_{1}-$1$_{0}$) line observed 
with the NRO 45~m telescope \citep[thin line; ][]{hirota2009} and the VLA (bold line). 
The VLA spectrum is obtained by integrating the channel map 
with the AIPS task ISPEC. 
\label{fig-sp}  }
\end{minipage}
\end{figure}

Figure \ref{fig-map} shows the integrated intensity map of the CCS ($J_{N}$=2$_{1}-$1$_{0}$) 
line toward L483, superposed on that of the CCS ($J_{N}$=4$_{3}-$3$_{2}$) line at 
the 45~GHz band taken with the NRO 45~m telescope \citep{hirota2009}. 
Hereafter we just refer to the latter map as "the single-dish CCS map", although 
the observed transition is different from that of the VLA observation. 
The shortest projected baseline in the VLA observation was 
1.85~k$\lambda$, and was comparable to the primary beam size of the VLA. 
Thus, the structure larger than about 60\arcsec \ or a linear size of 
12000~AU was significantly filtered out. 
Because we only observed the CCS ($J_{N}$=2$_{1}-$1$_{0}$) spectrum toward the 
core center with the NRO 45~m telescope, as shown in Figure \ref{fig-sp} \citep{hirota2009}, 
a combined image by using both the interferometer and the single-dish data 
\citep[e.g.][]{velusamy1995} could not be produced. 

As depicted in Figure \ref{fig-map}, the distribution of CCS consists of 
two compact clumps; one is located at 
northwest, (-28\arcsec, +12\arcsec) in right ascension and declination, 
respectively, of the protostar  IRAS~18148-0440, and the other is 
a weak component located from 
southeast to south of the protostar. The northwest clump has an FWHM 
size of 30\arcsec$\times$18\arcsec \ with the position angle of -26$^{\circ}$. 
Although the size is much smaller than the ridge structure extending from 
northwest to southeast seen in the single-dish CCS map \citep{hirota2009}, 
these compact clumps apparently trace its portions (Figure \ref{fig-map}). 
The southern clump agrees well with the local peak position of the single-dish map, 
which is slightly shifted to the south from the protostar position. 

The most remarkable feature found in the VLA map is 
that the distribution of CCS is clearly depleted at the protostar position. 
The radio continuum source, {\it{IRAS}} point source (IRAS~18148-0440), and dust continuum 
peaks \citep{beltran2001, fuller2000, jorgensen2004} all lie between the two CCS clumps. 
This central depletion corresponds to that previously reported in the NH$_{3}$ 
map taken with the VLA \citep{fuller2000}. 
Such a CCS hole is similar to the case of another protostellar core 
B335 \citep{velusamy1995} and a dynamically evolved prestellar core L1544 \citep{ohashi1999}, 
observed with VLA and BIMA, respectively. 
It is unlikely that this hole is due to the self-absorption, because the CCS ($J_{N}$=2$_{1}-$1$_{0}$) 
line is usually optically thin \citep{suzuki1992}. 
The hole can be interpreted as a consequence of real abundance decrease of CCS in 
the central part of a chemically and dynamically evolved core \citep{bergin1997, aikawa2001}. 
No signature of depletion in the single-dish CCS map of L483 \citep{hirota2009} 
originates from an insufficient spatial resolution to resolve the central hole. 

Figure \ref{fig-sp} presents the total flux density of the CCS ($J_{N}$=2$_{1}-$1$_{0}$) 
line observed with the NRO 45~m telescope whose beam size is 74\arcsec \ at 22~GHz 
\citep{hirota2009}, and the integrated flux density observed with the VLA. 
The VLA spectrum is obtained by summing over the emission region in the map by 
using the AIPS task ISPEC. 
The peak antenna temperature of the CCS ($J_{N}$=2$_{1}-$1$_{0}$) line is derived from 
the NRO 45~m telescope to be 0.36~K \citep{hirota2009}, 
corresponding to the total flux density of 1.18~Jy. 
On the other hand, the integrated flux density of the VLA map only recovers 0.43~Jy at the peak 
velocity channel. 
Even though we employ only a central part of the UV data, 
more than 60\% of the flux is missing in the VLA image. 
The peak velocity channel derived from the VLA map, 5.1~km~s$^{-1}$, 
is slightly blueshifted from that of the single-dish spectrum, 
5.36~km~s$^{-1}$ \citep{hirota2009}. 
A possible reason is that the systematic and redshifted velocity components 
are more extended than the blueshifted one, and are resolved out more 
significantly. 
This means that the CCS distribution does not have a compact central condensation. 

\begin{figure}[htb]
\epsscale{0.95}
\plotone{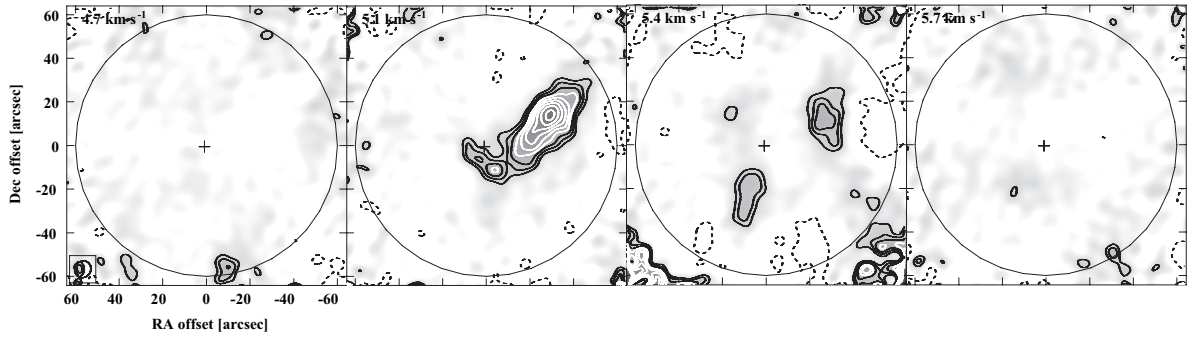}
\figcaption[]{Channel maps of the CCS ($J_{N}$=2$_{1}-$1$_{0}$) line observed 
with the VLA. 
A cross indicates the radio counterpart of IRAS~18148-0440 
\citep[VLA-2;][]{beltran2001}. The interval of the contours 
is 4~mJy~beam$^{-1}$ (1$\sigma$) and the lowest one is 12~mJy~beam$^{-1}$ (3$\sigma$). 
A dashed line represents a negative contour of -12~mJy~beam$^{-1}$. 
A circle indicates the primary beam size of the VLA. 
\label{fig-chmap}  }
\end{figure}

Figure \ref{fig-chmap} shows the channel maps of the CCS ($J_{N}$=2$_{1}-$1$_{0}$) line. 
We detected the CCS line only at the central two channels above the 3$\sigma$ noise 
level. However, we can see a weak emission feature around the southeast of 
the protostar position at the 5.7~km~s$^{-1}$ channel. 
This emission feature is also evident at the 5.4 and 
5.1~km~s$^{-1}$ channels, and hence, it could be real. 
Because of our coarse spectral resolution, $\sim$0.3~km~s$^{-1}$, 
it is difficult to detect a small velocity structures of the L483 core caused 
by rotation, infall, expansion, and/or outflow 
\citep[e.g.][]{myers1995, park2000, fuller2000, jorgensen2004}. 
In fact, no significant velocity gradient is seen between 
the northwest and southeast clumps. 

\section{Discussions}

Various molecular lines have been mapped 
in high-resolution with interferometers toward L483, 
including HCO$^{+}$, C$_{3}$H$_{2}$ \citep{park2000}, NH$_{3}$ \citep{fuller2000}, 
CN, C$^{18}$O, CS, HCN, HCO$^{+}$, N$_{2}$H$^{+}$, and their isotopic species 
\citep{jorgensen2004}. Single-dish mapping observations have also been 
conducted for some molecular lines \citep[e.g.][]{tafalla2000, hirota2009} and 
dust continuum emission \citep{fuller2000, jorgensen2004}. 
Nevertheless, none of them show a similar distribution to that of the CCS map taken 
with the VLA at first glance, except for the single-dish CCS map \citep{hirota2009}. 
Since the CCS linewidth is only less than 0.6~km~s$^{-1}$ and 
no correspondence with the velocity structure relevant to the outflow can be seen, 
the CCS clumps do not seem to interact with the outflow in contrast to 
the other molecular species \citep{jorgensen2004}. 

It is reported that the distribution of CCS and NH$_{3}$/N$_{2}$H$^{+}$ are 
anticorrelated with each other \citep{ohashi1999}.  This seems true for L483 as a whole. 
The NH$_{3}$ distribution appears to extend toward three directions, 
northwest, northeast, and southwest \citep{fuller2000}.  The northwestern clump of CCS is 
located at about 10\arcsec \ west of the peak of the northwest NH$_{3}$ ridge, 
although they are partly overlapped with each other.  Similarly, the distribution of CCS 
and N$_{2}$H$^{+}$ show mutual anticorrelation except for slight overlapping 
at the western and southeastern ends of the N$_{2}$H$^{+}$ and CCS distributions, 
respectively \citep{jorgensen2004}.  Interestingly, all of CCS, NH$_{3}$, and N$_{2}$H$^{+}$ 
show possible signature of depletion toward the protostar position 
\citep{fuller2000, jorgensen2004}. 

In contrast, C$^{18}$O and CS are enhanced in the vicinity of the protostar 
\citep{jorgensen2004}. This is interpreted as a consequence of evaporation from dust 
grains at the temperature above 40~K. In such a region, CH$_{4}$, whose sublimation 
temperature is 30~K, could also be evaporated from dust grains as expected in WCCC 
\citep{sakai2008, hassel2008}. 

It is surprising that the distributions of CCS and CS show anti-correlation, 
because CCS is believed to be closely related to CS 
\citep[e.g.][]{bergin1997, aikawa2001}. 
Our results indicate that CCS is hardly enhanced by the activity of the central protostar, 
namely, a compact centrally condensed structure that is related to WCCC is not 
detected. The observed CCS clumps might represent remnants of 
a quiescent cold outer envelope that still remains chemically young. 
This is consistent with the results of the mapping observation of C$_{4}$H toward 
L1527 and IRAS~15398$-$3359 \citep{sakai2008, sakai2009}, reporting that the significant 
fraction of C$_{4}$H is also distributed in the cold outer envelope as well as the vicinity 
of the protostar \citep{sakai2008, sakai2009}. 
The contribution from the cold outer envelope relative to that from the central 
condensation seems to be larger than the C$_{4}$H case in L1527. 

If above considerations are true, higher abundance of CCS in 
L483 \citep{hirota2009} implies its earlier evolutionary phase or 
faster contraction timescale than in the case of L1527 \citep{sakai2009}. 
According to the previous survey \citep{hirota2009}, the column densities of CCS are 
12.3$\times$10$^{12}$~cm$^{-2}$ and 5.1$\times$10$^{12}$~cm$^{-2}$ toward 
L483 and L1527, respectively. 
Although the NH$_{3}$/CCS ratio in L483 (121) is larger 
than that in L1527 (98) only by a factor of 1.2 \citep{hirota2009}, 
the column density of CCS in L483 is larger than L1527 by a factor of 2.4. 
The column density of CCS in IRAS~15398$-$3359 is not as high as that in 
L483 (N. Sakai, in preparation). On the other hand, 
the column densities of C$_{4}$H in L1527 and IRAS~15398$-$3359 are larger by 
a factor of about 2 than that in L483 \citep{sakai2009}. 
This is probably due to the heavier depletion of sulfur in L1527 and IRAS~15398$-$3359. 
The chemical model calculation by \citet{hassel2008} predicts that CCS is not very abundant in 
the WCCC environment compared with the cold early phase before onset 
of star-formation. 
Thus, the higher abundance of CCS in L483 means that its outer envelope would be in 
a less evolved phase, or close to the "cold peak" of chemical evolution \citep{hassel2008}. 
The chemical evolutionary phase of L483 would be rather close to those in typical 
prestellar cores such as L1544 \citep{ohashi1999, aikawa2001} or chemically young 
dark cloud cores which are rich in carbon-chain molecules \citep{hirota2006}. 
It is also likely that such a difference can be an evidence of variation of contraction 
and depletion timescale between L483 and other WCCC sources. 
A similar idea has been proposed for a starless carbon-chain rich core L492, 
which is also located in the Aquila rift \citep{hirota2006}, as well as for 
recent survey observations of carbon-chain molecules \citep{sakai2009, hirota2009}. 

It should be noted that the anti-correlation between CCS and CS could be 
attributed to the difference in the gas densities traced by these lines. 
The critical densities for the CCS ($J_{N}$=2$_{1}-$1$_{0}$) and CS ($J$=2$-$1) lines 
are 8$\times$10$^{4}$~cm$^{-3}$ and 1.7$\times$10$^{6}$~cm$^{-3}$, respectively. 
Hence, the VLA map of the CCS ($J_{N}$=2$_{1}-$1$_{0}$) line 
would preferentially trace the outer less-dense regions. 

In order to address the issue whether CCS is not really 
related to the WCCC possibly occurring in deep inside the dense and warm 
gas in L483, observations of CCS in higher frequency bands would be required. 
For example, the critical densities of the CCS ($J_{N}$=7$_{6}-$6$_{5}$, 81~GHz) 
and CCS ($J_{N}$=8$_{7}-$7$_{6}$, 93~GHz) lines are 
4.2$\times$10$^{6}$~cm$^{-3}$ and 
6.9$\times$10$^{6}$~cm$^{-3}$, respectively, 
which can trace almost the same volume of gas as of the CS ($J$=2$-$1) line. 
Future observations with ALMA will be able to reveal the high-resolution 
CCS distributions at the vicinity of the newly born protostar IRAS~18148-0440, 
along with the extended outer envelope, thanks to its high dynamic-range 
imaging capability. 
In addition, it would be essential to investigate the distributions of CCS 
in other prototypical WCCC sources, L1527 and IRAS~15398$-$3359, 
with high resolution observations in order to understand the production 
mechanism of CCS in the WCCC environment. 

Detailed studies on WCCC in comparison with hot corino chemistry 
are important to understand chemical variation in 
low-mass star-formation processes followed by planetary formation \citep{sakai2009}. 
Thus, further high resolution observations, in particular with ALMA, 
of variety of longer carbon-chain molecular species including 
C$_{n}$H and C$_{n}$H$_{2}$, HC$_{n}$N and C$_{n}$H$^{-}$, 
as well as C$_{n}$S, will be crucial for L483 and more protostellar 
cores in order to shed light on the chemical 
and dynamical evolutionary scheme of low-mass star-formation.

\acknowledgements

We are grateful to the staff of the NRAO, 
in particular Claire J. Chandler and Mark J. Claussen, for support of our project. 
The NRAO is a facility of the National Science Foundation operated under cooperative 
agreement by Associated Universities, Inc. 
This study is supported by Grain-in-Aids from Ministry of Education, 
Culture, Sports, Science, and Technologies (21224002 and 21740132). 

{\it Facilities:} \facility{VLA}.

{}

\end{document}